\documentclass[a4paper]{jpconf} 
\usepackage{graphicx}

\usepackage{lineno}

\modulolinenumbers[1]

\begin{document}
\title{Deferred Optical Photon simulation for the JUNO experiment}

\author{Tao Lin (on behalf of the JUNO collaboration)}

\address{Institute of High Energy Physics, Chinese Academy of Sciences, Beijing 100049, China}

\ead{lintao@ihep.ac.cn}

%

\begin{abstract}

The Jiangmen Underground Neutrino Observatory (JUNO) is designed to determine the neutrino mass ordering and precisely measure oscillation parameters. It is being built in South China at a depth of 700~m underground and comprises a central detector, water Cerenkov detector and top tracker. The central detector is designed to detect anti-neutrinos with an energy resolution of 3\% at 1~MeV, using a 20 kt liquid scintillator target with 17,612 20-inch PMTs and 25,600 3-inch PMTs. The scintillator provides a light yield of approximately 10,000 photons per MeV. Monte Carlo simulation is a crucial tool for developing an understanding of detector performance, requiring the production of large samples of background processes with optical photons. Simulation of large numbers of optical photons with Geant4 is computationally challenging for both processing time and memory resources. In order to optimize resource usage, a deferred optical photon simulation workflow is proposed and implemented using Geant4 classes. The key idea is to simulate events initially without optical photons, only performing the optical photon simulation when user specified criteria are met.

In this contribution, the design and the implementation of the deferred optical photon simulation will be presented. Optical simulation comprises generation of photons and propagation through the detector implementing optical physics processes including reflection, refraction, scattering and absorption. Instead of generating the optical photons at each step by Geant4 immediately, the necessary data to generate optical photons at each step are collected, which is termed {\tt GenStep}. At the end of each event, user specified criteria determines if the optical photon simulation is performed using a class called {\tt G4OPSimulator}. The class {\tt G4OPSimulator} implements a customized simulation workflow, based on Geant4 internal classes including {\tt G4TrackingManager} and {\tt G4StackManager}. The simulator passes references to the collected {\tt GenStep} objects to customized Scintillation and Cerenkov processes which generate the optical photons in {\tt G4Track} objects. As a track could be absorbed and re-emitted, the secondaries will be retrieved from the {\tt G4TrackingManager} and pushed to the {\tt G4StackManager}. The performance of the simulator will be presented. The technique of deferred optical photon simulation can be applied to all event types and it is expected to be particularly beneficial with rare processes. Especially the events must be selected during the detector simulation at runtime, instead of the event generation. An application of the technique to the simulation of such events will be shown.
\end{abstract}

\section{Introduction}
The Jiangmen Underground Neutrino Observatory (JUNO) experiment under construction in southern China, will have a rich physics program~\cite{An:2015jdp,JUNO:2022hxd} including neutrino mass ordering determination and a precise measurement of three oscillation parameters. With 20~kt liquid scintillator (LS) and an energy resolution of 3\% at 1~MeV~\cite{JUNO:2020xtj}, the JUNO detector could detect reactor $\nu$, supernova $\nu$, geo-$\nu$, atmospheric $\nu$, solar $\nu$ etc. Anti-neutrinos are detected via the inverse beta decay (IBD) process ($\bar{\nu}_e + p \to e^{+} + n$). The positrons and neutrons deposit energy in the LS, then the optical photons are produced via scintillation process and Cerenkov process. The optical photons transport in the detector and some of them are absorbed. Finally, the optical photons are detected by the photomultiplier tubes (PMTs) and converted to photoelectrons (PEs). 

\begin{figure}[h]
    \begin{center}
    \includegraphics[width=0.65\linewidth]{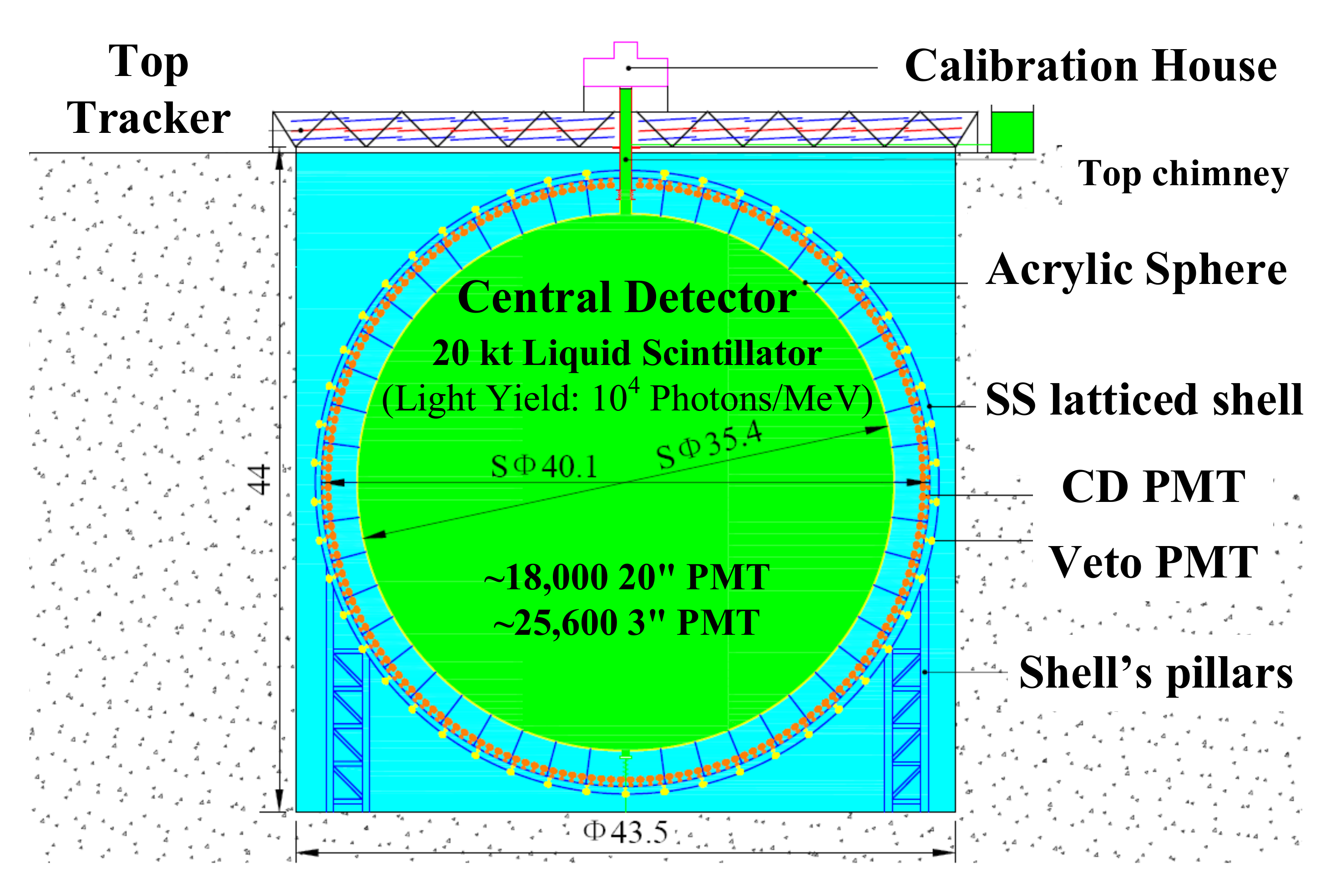}
    \end{center}
    \caption{Schematic view of the JUNO detector\label{fig:detector}}
\end{figure}

As shown in figure~\ref{fig:detector}, the innermost part of the JUNO detector is a central detector (CD) containing LS and equipped with 17,612 20~inch PMTs and 25,600 3~inch PMTs. Water is filled around the CD to reduce the backgrounds from natural radioactivity. The outer water pool is also used to veto the cosmic ray muons by detecting Cerenkov photons with the veto PMTs. On the top of the water pool, a top tracker is used to measure muons. 

\section{Motivation}
Monte Carlo simulation is a crucial tool for developing an understanding of the detector performance. The JUNO simulation software consists of physics generators, detector simulation and electronics simulation. The physics generators produce the primary track information in HepMC format~\cite{Dobbs:2001ck}. Then the HepMC objects are converted to track objects in Geant4 and the track objects are passed to the detector simulation. The detector simulation simulates the detector response and produces the Monte Carlo truth information, including the track objects and hit objects. Then the hit objects are read by the electronics simulation and the waveforms are generated by the electronics simulation. 

The JUNO detector simulation software~\cite{Lin:2017usg} is developed based on Geant4~\cite{GEANT4:2002zbu,Allison:2006ve} with precise optical processes, including scattering, absorption, re-emission, reflection and refraction. Due to the large numbers of optical photons, the optical processes are the most time consuming part in the simulation, especially for the high energy cosmic ray muons which will produce millions of optical photons. Several methods have been already developed to speed up the simulation of high energy events, such as Opticks~\cite{Blyth:2021gam}, voxel method~\cite{Lin:2016vua}, machine learning based method etc. These methods will collect the necessary information and then replace the Geant4 simulation of optical photons. Meanwhile, simulation of large samples of low energy backgrounds with optical photons is also time consuming. 

However, background events with optical photon simulation are not always needed. For example, only part of them could reach the LS and generate scintillation photons. Most of the external radioactivity backgrounds could not reach the LS. However, photons could be still generated via Cerenkov process and then simulated by Geant4. As such events could be lower than the detector trigger threshold, they would be finally abandoned in the later data processing. In order to only perform optical photon simulation when selection criteria are met, a deferred optical simulation workflow is proposed and implemented based on the Geant4 simulation framework.

\section{Design and Implementation}


In Geant4, a particle (or a {\tt G4Track} object) is simulated by the {\tt G4TrackingManager}. A track is divided into multiple steps and each step is simulated by {\tt G4SteppingManager}. Optical photons are generated at each step immediately and the corresponding {\tt G4Track} objects are pushed into the {\tt G4StackManager} object. The simulation of original particle will be paused and these optical photons will be popped from {\tt G4StackManager} and then simulated by {\tt G4TrackingManager}. Only after these optical photons are processed, the simulation of the original particle will be continued. The key idea of deferred optical photon simulation is to simulate events initially without optical photon, only performing the optical photon simulation when user-specified criteria are met. 

\begin{figure}[h]
    \begin{center}
    \includegraphics[width=0.5\linewidth]{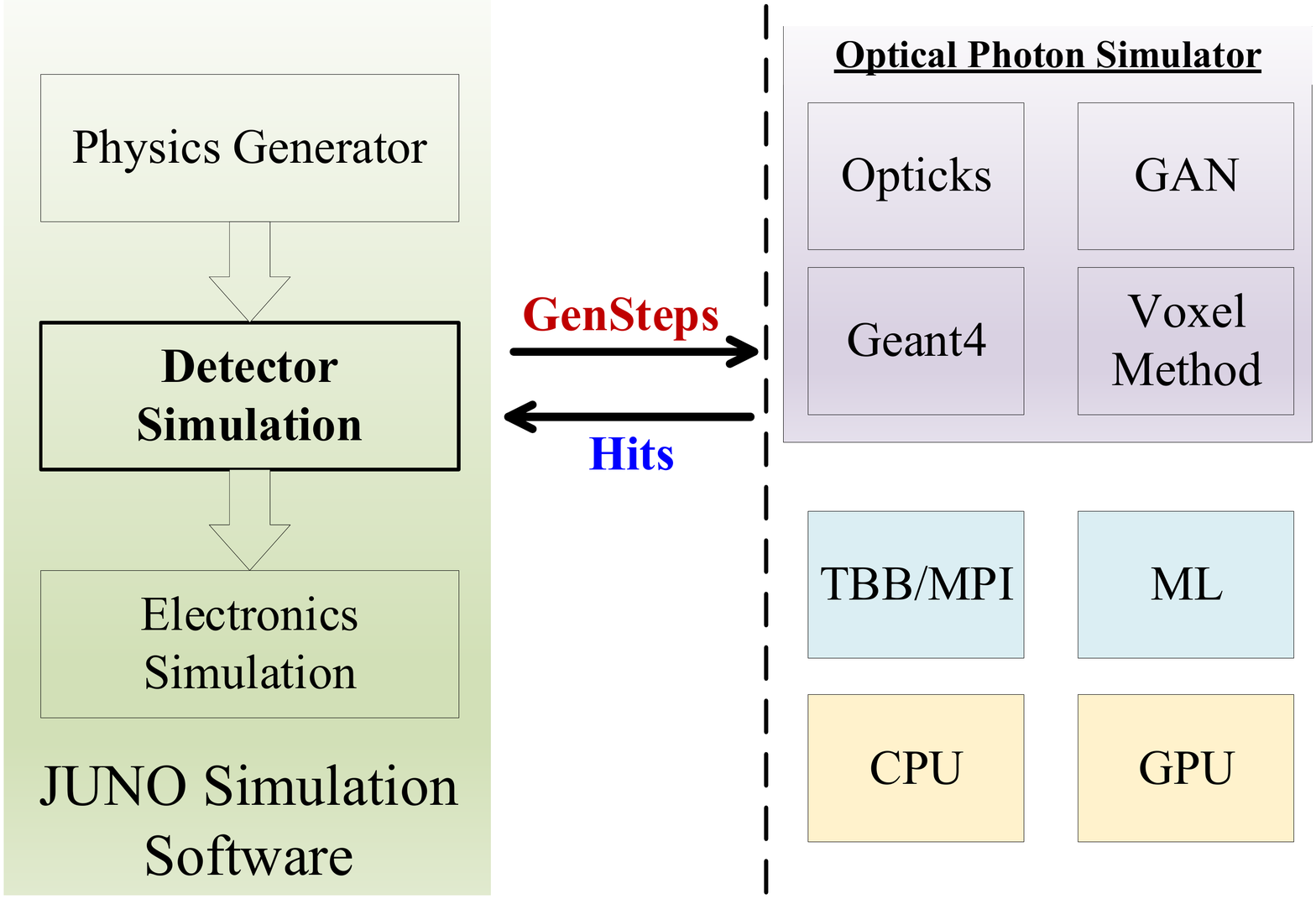}
    \end{center}
    \caption{The unified interface for different optical photon simulators\label{fig:design}}
\end{figure}

Instead of collecting the generated optical photons, the information (so called {\tt GenStep} in Opticks) used to generate optical photons at each step will be collected first. If the selection criteria is met, then these information will be used to generate and simulate optical photons. As shown in figure~\ref{fig:design}, {\tt GenStep} objects could be passed to the different optical photon simulators and then the simulated hit objects are returned to the Geant4 simulation. In this design, the simulation of optical photons could be offloaded to the different simulators. These simulators could be then implemented using different techniques, while the interfaces to these simulators are kept same. 

\begin{figure}[h]
    \begin{center}
    \includegraphics[width=0.65\linewidth]{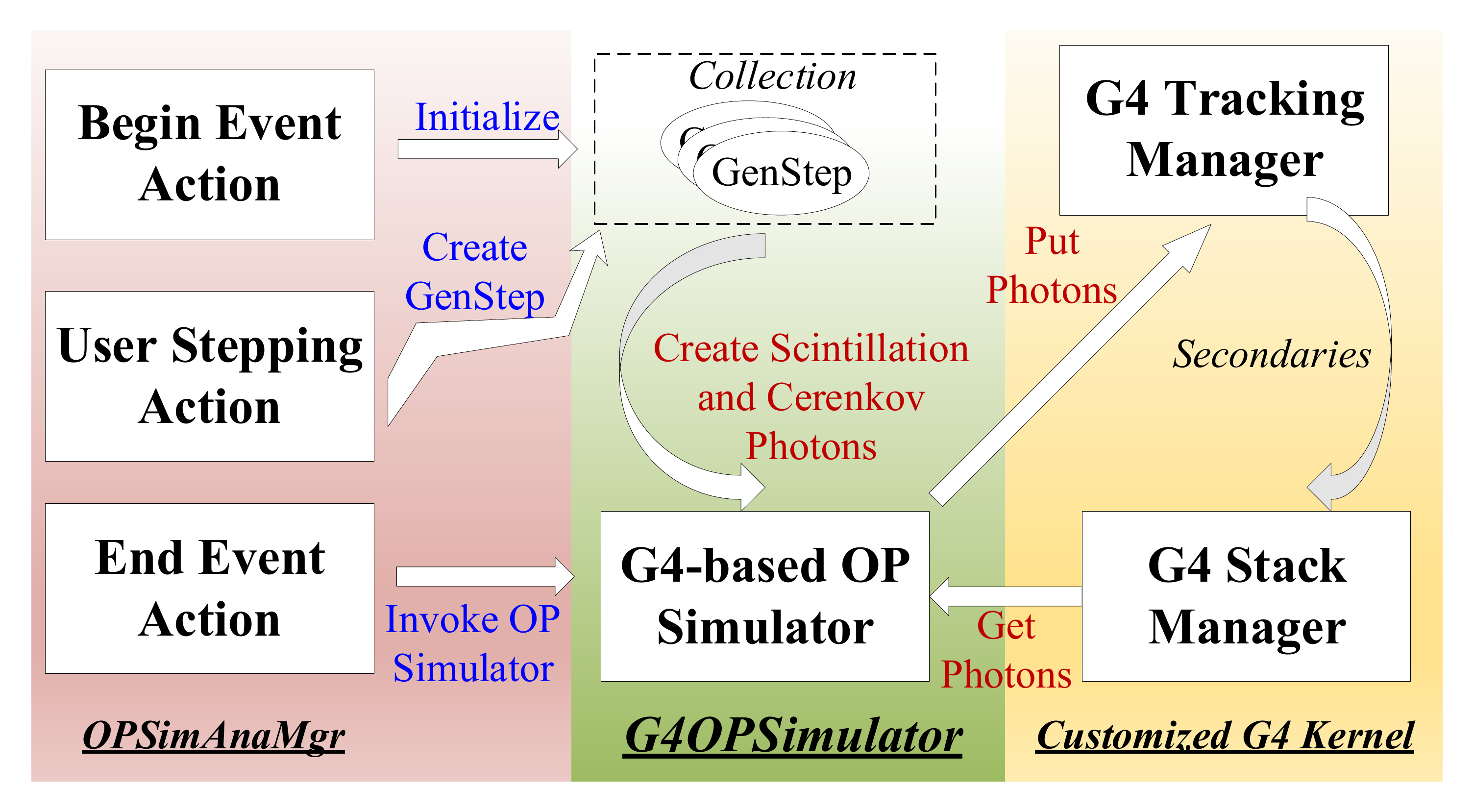}
    \end{center}
    \caption{Geant4 based Optical Photon simulator\label{fig:g4op}}
\end{figure}

In order to defer the simulation of optical photons in Geant4, a new simulator {\tt G4OPSimulator} has been developed. Figure~\ref{fig:g4op} shows the implementation of the Geant4 based optical photon simulator. It consists of a Geant4 user action, the {\tt G4OPSimulator} and a customized Geant4 kernel. The Geant4 user action provides the user interface to retrieve necessary information from Geant4. The {\tt G4OPSimulator} controls the generation and simulation of optical photons. The customized Geant4 kernel is invoked by the {\tt G4OPSimulator} and provides the optical photon simulation based on Geant4 internal classes. 

When simulating a new event, a Geant4 user action called {\tt OPSimAnaMgr} is used to fill the collection of {\tt GenStep} objects and invoke the {\tt G4OPSimulator}. At the beginning of this event, it will reset the collection of {\tt GenStep} objects. Then at each step, a {\tt GenStep} object is created and pushed to the collection. At the end of this event, it will invoke the optical photon simulator if the selection criteria are met. The selection criteria could be implemented by the users. For example, if the total deposit energy in the LS is lower than a threshold, then the optical photon simulator will not be invoked. For those events which must be selected at runtime instead of generation, this method is useful. 

When the {\tt G4OPSimulator} is invoked by {\tt OPSimAnaMgr}, it will initialize all the necessary parameters for generation of scintillation photons and Cerenkov photons. These parameters are retrieved from the corresponding Geant4 physics processes, so that the necessary parameters could be kept consistent. Then, a {\tt GenStep} object is retrieved from the collection and optical photons are created from the {\tt GenStep} object. As {\tt GenStep} is not an official class in Geant4, the conversion from the {\tt GenStep} object to the {\tt G4Track} objects is not automatically performed. Therefore, the generation of scintillation photons and Cerenkov photons are re-implemented in the {\tt G4OPSimulator}. Instead of generating the optical photons at once, the optical photon is generated and processed by the {\tt G4TrackingManager} one by one. Optical photon could be absorbed and re-emit again, so a new {\tt G4Track} will be created and stored in {\tt G4StackManager}. Therefore,  {\tt G4OPSimulator} will try to pop a track from this stack. In this manner, the memory usage  could be reduced during simulation.

Even though the workflow of simulation is changed, Geant4 sensitive detectors still works. When a photon is propagated to a PMT, the associated sensitive detector will be invoked and a hit object will be created. All the created hits will be stored in the hit collection, so these hit objects could be retrieved at the end of event. For the events without optical photon simulation, the hit collection would be empty.

\section{Performances}
The simulation performances are measured using 1 MeV gamma at the detector center. For each test, 1000 events are simulated and the simulation time is saved. The initialization is not included in the measured time. A dedicated computing node is used to run all the tests. There are two rounds of measurements. In the first round, the deferred simulation is enabled and all the 1000 events are simulated with optical photons. Then the total numbers of the collected hits are compared between the default simulation and the deferred simulation. In the second round, the deferred simulation is still enabled, however, only a part of the events are simulated with optical photons while the others are simulated without optical photons. In this dummy test, an extra parameter ``ratio'' is used to control the number events to simulate with optical photons.

\begin{figure}[h]
    \begin{center}
    \includegraphics[width=0.45\linewidth]{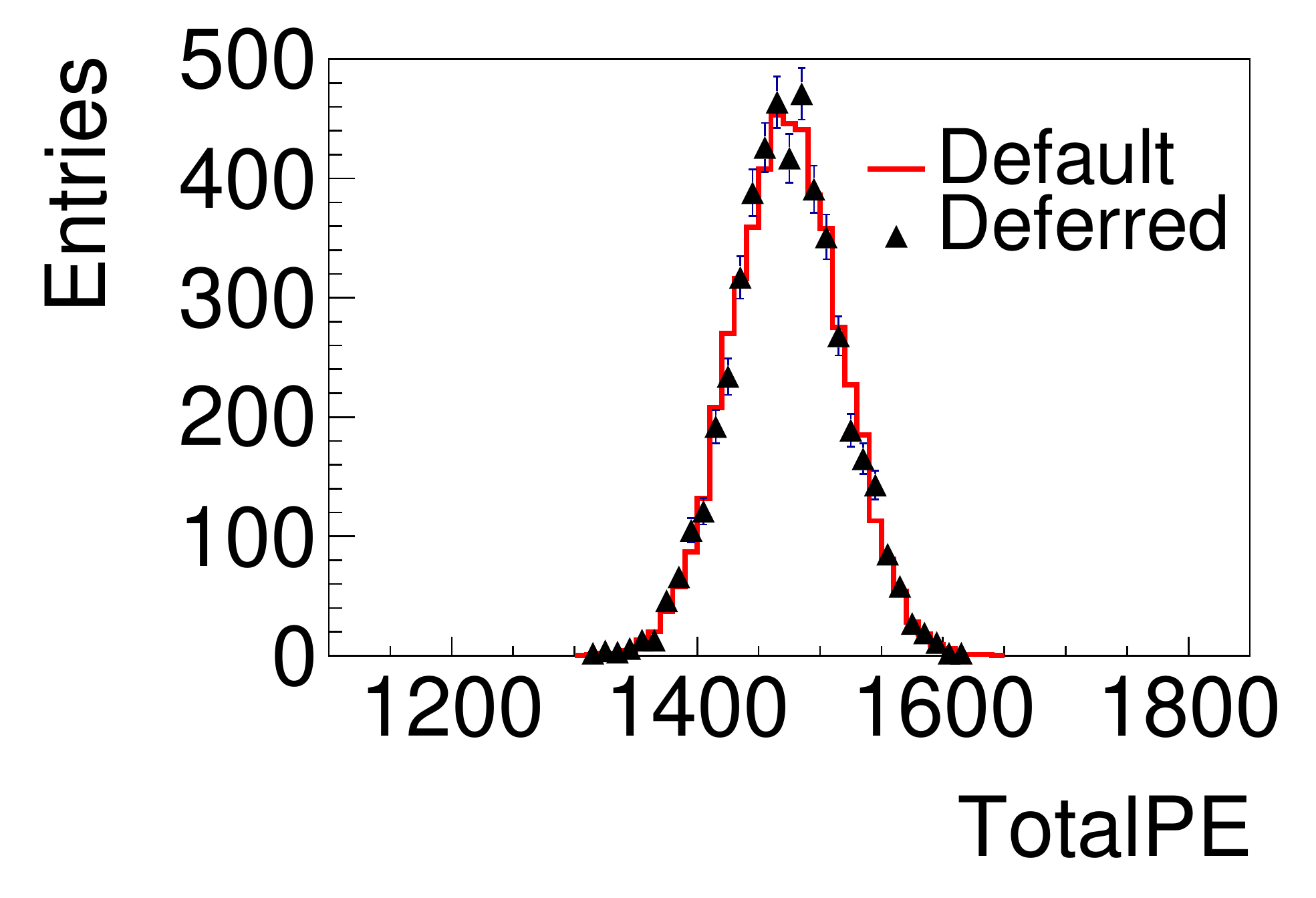}
    \includegraphics[width=0.45\linewidth]{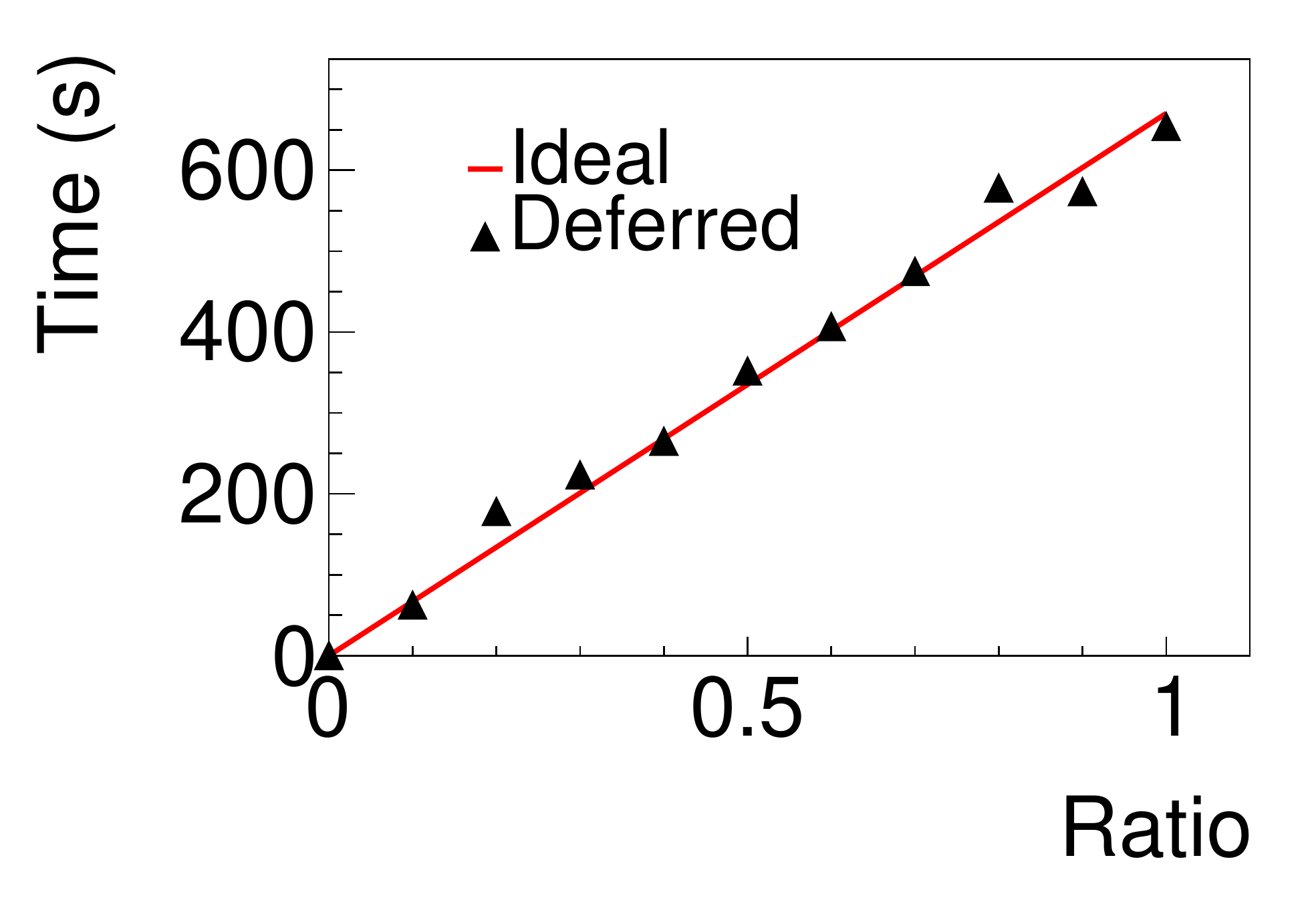}
    \end{center}
    \caption{The performances of the optical photon simulator \label{fig:result}}
\end{figure}

The measured performances are shown in figure~\ref{fig:result}. The left figure shows the results using deferred and default simulation. Due to the simulation workflow is changed, the sequences of random numbers are not exactly same for the deferred and default simulation, so only the statistical results are compared. The results show that the deferred simulation is correct.

The right figure shows the simulation time versus the number of simulated events with optical photon simulation. The ratio in the figure is defined as N(events with OP) / N(all events). The red ideal line is based on the ratio $\times$ T(default), while the ideal speed-up is 1/ratio. By reducing the number of events with optical photon simulation, the method could improve the speed. 

\section{Conclusions}
Simulating a large number of events is time consuming, especially with optical processes. In this paper, a deferred optical photon simulation is presented. By applying the deferred simulation and selection criteria, the total simulation time could be reduced while the interested events are all kept. Currently, this method is already adopted in the simulation of several background samples. This method will be enabled by default and used for the Monte Carlo production in the future.

\ack
This work is supported by National Natural Science Foundation of China (12025502, 11805223), the Strategic Priority Research Program of the Chinese Academy of Sciences (Grant No. XDA10010900), the Innovation Project of the Institute of High Energy Physics, and Xie Jialing Fund.

\section*{References}
\bibliography{iopart-num}

\end{document}